\documentclass[aps,superscriptaddress,twocolumn]{revtex4}%
\usepackage{mathrsfs}
\usepackage{amsfonts}
\usepackage{amsmath}
\usepackage{amssymb}
\usepackage{graphicx}
\usepackage{diagbox}
\usepackage{booktabs}
\usepackage{epstopdf}
\usepackage{graphicx}
\usepackage{mathrsfs}
\usepackage{bm}
\usepackage{dcolumn}
\usepackage{epstopdf}
\usepackage{dsfont}
\usepackage{tabularx}
\usepackage{float}
\usepackage{color}
\usepackage{pifont}
\usepackage{siunitx}
\usepackage{multirow}%
\providecommand{\U}[1]{\protect\rule{.1in}{.1in}}

\begin{document}
\title{Tunable Fano and Dicke effects in quantum transport of double quantum dots
sandwiched between topological insulators}
\author{Yuan Hong}
\affiliation{Institute of Applied Physics and Computational Mathematics and National Key
Laboratory of Computational Physics, Beijing 100088, China}
\affiliation{Graduate School, China Academy of Engineering Physics, Beijing 100088, China}
\author{Zhen-Guo Fu}
\thanks{Corresponding author. Email address: fu\_zhenguo@iapcm.ac.cn}
\affiliation{Institute of Applied Physics and Computational Mathematics and National Key
Laboratory of Computational Physics, Beijing 100088, China}
\author{Zhou-Wei-Yu Chen}
\affiliation{School of Mathematics and Computational Science, Xiangtan University,
Xiangtan, 411105, China}
\author{Feng Chi}
\affiliation{School of Electronic and Information Engineering, Zhongshan Institute,
University of Electronic Science and Technology of China, Zhongshan 528400, China}
\author{Zhigang Wang}
\affiliation{Institute of Applied Physics and Computational Mathematics and National Key
Laboratory of Computational Physics, Beijing 100088, China}
\author{Wei Zhang}
\thanks{Corresponding author. Email address: zhang\_wei@iapcm.ac.cn}
\affiliation{Institute of Applied Physics and Computational Mathematics and National Key
Laboratory of Computational Physics, Beijing 100088, China}
\author{Ping Zhang}
\thanks{Corresponding author. Email address: zhang\_ping@iapcm.ac.cn}
\affiliation{Institute of Applied Physics and Computational Mathematics and National Key
Laboratory of Computational Physics, Beijing 100088, China}

\begin{abstract}
We study the quantum transport in double quantum dots (DQD) sandwiched between
surfaces of topological insulator (TI) Bi$_{2}$Te$_{3}$, which possess strong
spin-orbit coupling (SOC) and $^{d}$C$_{3v}$ double group symmetry. Different
from the spin-conserved case with two-dimensional electron gas (2DEG)
electrodes, the conductance displays a \textit{universal scaling relation} for
different Fermi energy associated with the topological nature/linear
dispersion of topological surface states. The interplay between direct
inter-dot tunneling and surface state mediated interaction leads to tunable
Dicke and Fano effects with changing the inter-dot distance. We propose
nano-rulers with different measurement range and resolution based on the Fano
$q$-factor. Furthermore, when applying an in-plane Zeeman field, a crossover
from a double-peak shape to a quad-peak shape in conductance curve appears.
Moreover, the rotational symmetry of the system could also be revealed from
the conductance pattern. Our findings contribute to a better understanding of
the quantum transport in the presence of electrode's SOC topological states.

\end{abstract}
\maketitle


\emph{Introduction.}---Quantum dots (QDs) are widely studied due to their
excellent properties
\cite{QD_RMP,SpinQD1,SpinQD2,SpinQD3,QDcoherence1,QDcoherence2,QDcoherence3,WangYK2024}%
, thanks to their compact size. They could be arranged into superstructures
such as arrays or chains, which offer a key platform to realize novel quantum
coherence effects
\cite{Zhu1,Zhu2,MZMQD1,MZMQD2,MZMQD3,MZMQD4,matern2023metastability,majek2022spin,feng2023control}%
. One of the simplest realizations is a parallel coupled double quantum dots
(DQD) sandwiched between electrodes at both the right and left sides. As the
inter-dot separation $d$ becomes smaller than the system characteristic
length-scale, i.e., the electrode Fermi wavelength $k_{F}^{-1}$, the coherence
effect, arising from the electrode-mediated coupling between QDs, becomes
significant. For instance, a system consisting of two-dimensional electron gas
(2DEG) electrodes displays a conductance lineshape with a narrow peak
superposes on a much wider peak in the case of $d\mathtt{\cdot}k_{F}%
\mathtt{<}1$, which is believed to share the same mechanism as the Dicke
super-radiation
\cite{shahbazyan1994,shahbazyan2011,Dicke_Fano1,Dicke_Fano2,Dicke_Fano3,Dicke_Fano4}
hence is named after it.

The 2DEG-based study \cite{shahbazyan2015} also indicates that the exotic
lineshape is smeared out when considering the weak Rashba spin-orbit coupling
(SOC) in the electrode. Instead of treating the SOC perturbatively, we adopt
the strong SOC limit as our starting point and explore the associated
topological effect and interference effect on the quantum transport.
Therefore, we investigating the resonant tunneling in a DQD system sandwiched
between topological insulator surface electrodes
\cite{hasan,SCZhang,shen2012topological,3DTI,modelTI,moore2010birth}, such as
Bi$_{2}$Te$_{3}$ \cite{Zhang2009,Bi2Te3_1,Bi2Te3_2}. From the perspective
symmetry, key features such as the height (DQD degeneracy) and interference of
resonance peaks are systematically studied.

In the absence of magnetic field, we uncover a universal scaling relation of
the conductance for different Fermi energy. The conductance exhibits tunable
Fano and Dicke effects due to the interplay between direct inter-dot tunneling
and surface state mediated interaction. Additionally, the coherent effects
modulated by an in-plane Zeeman field have also been addressed.

\begin{figure*}[ptb]
\centering
\includegraphics[width=1\linewidth]{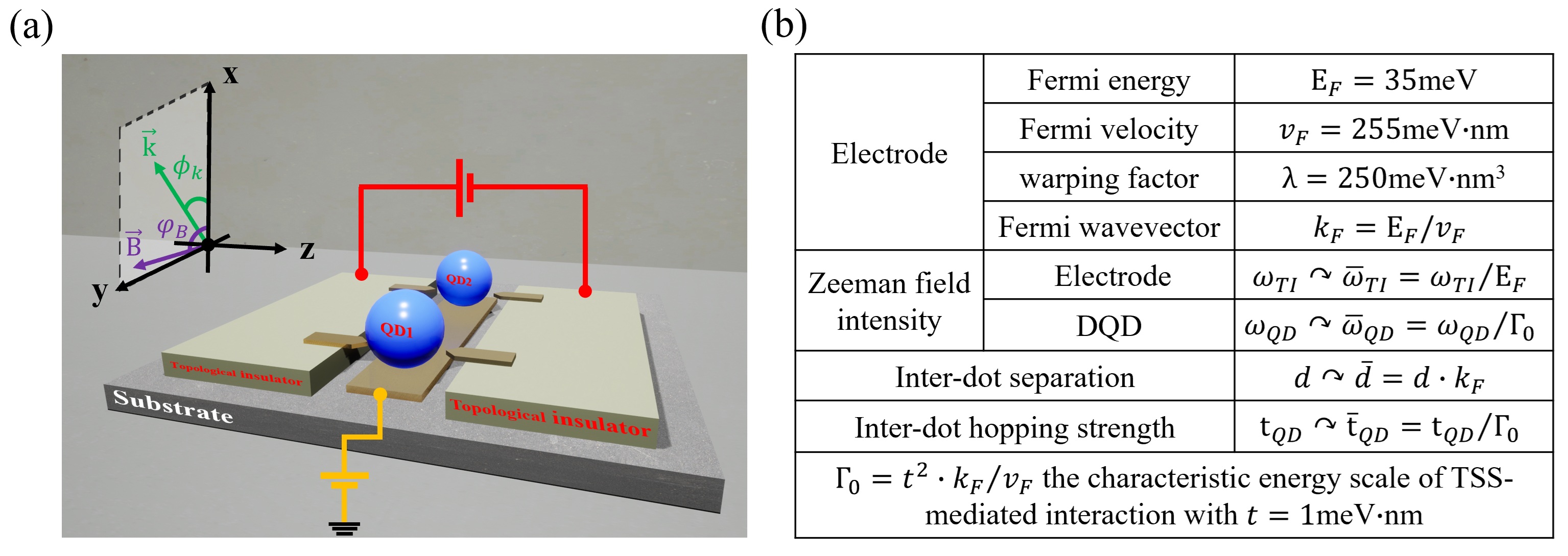} \caption{(Color online) (a)
Schematic of resonant tunneling in DQD sandwiched TSS electrodes with an
in-plane Zeeman field. (b) Parameters and dimensionless variables. }%
\label{fig1}%
\end{figure*}

\emph{Hamiltonian and Method.}---Our system with sandwiched structure consists
of a DQD and peripheral Bi$_{2}$Te$_{3}$ electrodes (as shown in
Fig.\thinspace\ref{fig1}(a)) and is described by the Hamiltonian
$H_{tot}\mathtt{=}H_{DQD}\mathtt{+}H_{TI}^{L}\mathtt{+}H_{TI}^{R}%
\mathtt{+}H_{int}$. Here, the DQD Hamiltonian is written as
\begin{align}
H_{DQD}=\sum\limits_{js}\sum\limits_{j^{\prime}s^{\prime}}  &  \left[
E_{0}\tau_{jj^{\prime}}^{0}\sigma_{ss^{\prime}}^{0}+t_{QD}\tau_{jj^{\prime}%
}^{x}\sigma_{ss^{\prime}}^{0}\right. \nonumber\\
&  \left.  +\omega_{QD}\tau_{jj^{\prime}}^{0}\left(  \boldsymbol{\sigma}%
\cdot\boldsymbol{n}_{B}\right)  _{ss^{\prime}}\right]  d_{js}^{\dagger
}d_{j^{\prime}s^{\prime}},
\end{align}
where $\sigma^{0}/\tau^{0}$ and $\sigma^{i}/\tau^{i}$ ($i\mathtt{=}x,y,z$)
respectively being the identity and Pauli matrices in spin and DQD
configuration spaces, with $\boldsymbol{\sigma}$$\mathtt{=}\left(  \sigma
^{x},\sigma^{y},\sigma^{z}\right)  $ and $\boldsymbol{\tau}\mathtt{=}\left(
\tau^{x},\tau^{y},\tau^{z}\right)  $. $\omega_{QD}$ and $\boldsymbol{n}_{B}%
$=$(\cos\phi_{B},\sin\phi_{B},0)$ are the intensity and direction vector of
the in-plane Zeeman field, $E_{0}$ and $t_{QD}$ denote the resonant energy
level and inter-dot hopping strength of the DQD. The interaction between the
DQD and the electrodes is given by
\begin{equation}
H_{int}=\sum\limits_{\mathbf{k}\delta\alpha}\sum\limits_{js}\left(
V_{\boldsymbol{k}\delta\alpha}^{js}d_{js}^{\dagger}c_{\boldsymbol{k}%
\delta\alpha}+h.c.\right)  ,
\end{equation}
where the simplification $V_{\boldsymbol{k}\delta\alpha}^{js}\mathtt{\approx
}t\Psi_{\boldsymbol{k}\delta\alpha}^{s}\left(  \boldsymbol{r}_{j}\right)  $
\cite{shahbazyan1994} is adopted with $\boldsymbol{r}_{j}$ the QD coordinate
projection in the electrode plane.

The electrode located at $\alpha\mathtt{=}L,R$ side is described by the
topological surface state (TSS) Hamiltonian derived from the 2D surface band
structure of 3D topological insulator Bi$_{2}$Te$_{3}$ using the
$k\mathtt{\cdot}p$ method \cite{Bi2Te3_1}%

\begin{align}
H_{TI}^{\alpha}=\sum\limits_{\boldsymbol{k}ss^{\prime}}  &  \left[
v_{f}\left(  \boldsymbol{\sigma}_{ss^{\prime}}\times\boldsymbol{k}\right)
_{z}+\frac{\lambda}{2}\left(  k_{+}^{3}+k_{-}^{3}\right)  \sigma_{ss^{\prime}%
}^{z}\right. \nonumber\\
&  \left.  +\omega_{TI}\left(  \boldsymbol{\sigma}_{ss^{\prime}}%
\boldsymbol{\cdot n}_{B}\right)  \right]  c_{\boldsymbol{k}s\alpha}^{\dagger
}c_{\boldsymbol{k}s^{\prime}\alpha}, \label{H_TI}%
\end{align}
\newline Here, $\boldsymbol{k}\mathtt{=}k\mathtt{\cdot}\boldsymbol{n}_{k}$ is
the wavevector of the TSS, followed by $k_{\pm}\mathtt{=}ke^{\pm i\varphi_{k}%
}$ and $\boldsymbol{n}_{k}\mathtt{=}\left(  \cos\varphi_{k},\sin\varphi
_{k},0\right)  $ with $k$ and $\varphi_{k}$ the radius and angle of the
wavevector. $v_{f}$ and $\lambda$ the Fermi velocity and warping factor, and
$\omega_{TI}$ the Zeeman field intensity. Eq.\thinspace(\ref{H_TI}) could also
be re-formulated as a two-level system in an effective Zeeman field, namely
\[
H_{TI}^{\alpha}=\sum\limits_{\boldsymbol{k}ss^{\prime}}\left[
\boldsymbol{\sigma}\cdot\mathcal{B}\left(  \boldsymbol{k}\right)  \right]
_{ss^{\prime}}c_{\boldsymbol{k}s\alpha}^{\dagger}c_{\boldsymbol{k}s^{\prime
}\alpha}%
\]
with
\begin{align*}
\mathcal{B}_{x}\left(  \boldsymbol{k}\right)   &  =\omega_{TI}\cos\phi
_{B}+v_{f}k\sin\varphi_{k},\\
\mathcal{B}_{y}\left(  \boldsymbol{k}\right)   &  =\omega_{TI}\sin\phi
_{B}-v_{f}k\cos\varphi_{k},\\
\mathcal{B}_{z}\left(  \boldsymbol{k}\right)   &  =\lambda k^{3}\cos
3\varphi_{k}.
\end{align*}
By solving the secular equation, the energy dispersion and corresponding
chiral eigenstate are given by
\begin{align}
\Psi_{\boldsymbol{k}\delta\alpha}\left(  \boldsymbol{r}\right)   &
=e^{i\boldsymbol{k}\mathbf{\cdot}\boldsymbol{r}}%
\begin{pmatrix}
A_{\delta} & -i\delta A_{\bar{\delta}}e^{i\theta_{k}^{B}}%
\end{pmatrix}
^{T},\nonumber\\
E_{\boldsymbol{k}\delta\alpha}  &  =\delta\sqrt{\left\vert \xi_{k}%
^{B}\right\vert ^{2}+\lambda^{2}k^{6}\cos^{2}\left(  3\varphi_{k}\right)  }%
\end{align}
with $\bar{\delta}\mathtt{=-}\delta$ and $\delta\mathtt{=\pm}1$ being the
chirality, and
\begin{align}
A_{\delta}  &  =-\sqrt{\left[  \left\vert E_{\boldsymbol{k}\delta\alpha
}\right\vert +\delta\lambda k^{3}\cos\left(  3\varphi_{k}\right)  \right]
/2\left\vert E_{\boldsymbol{k}\delta\alpha}\right\vert },\nonumber\\
\xi_{k}^{B}  &  =|\xi_{k}^{B}|e^{i\theta_{k}^{B}}=v_{f}ke^{i\varphi_{k}%
}+i\omega_{TI}e^{i\phi_{B}}.
\end{align}
One can seen that both the wavefunction and the dispersion pose a $2\pi/3$
periodicity on the orientation of Zeeman field, which has important physical
consequence as shown later. In the absence of warping effect ($\lambda
\mathtt{=}0$), where the spin distribution is confined within the
$s_{x}\mathtt{-}s_{y}$ plane, the electrode is characterized by a nonzero
winding number around the singularity (defined by $\left\vert \mathcal{B}%
\left(  \boldsymbol{k}\right)  \right\vert \mathtt{=}0$, also known as the
Dirac point)%
\begin{equation}
w_{\delta}=\frac{1}{\pi}\int_{C}d\boldsymbol{k}\frac{\mathcal{B}_{x}\left(
\boldsymbol{k}\right)  \partial_{k}\mathcal{B}_{y}\left(  \boldsymbol{k}%
\right)  }{\left\vert \mathcal{B}\right\vert ^{2}}=-\delta,
\end{equation}
thus is classified as a Dirac semimetal \cite{dirac_semi,Zhai_2021}. In such
cases, due to the topological protection, applying an in-plane Zeeman field
could only shift the Dirac cone around the momentum space. Aside from the
response on in-plane Zeeman field, another distinguishing feature of the Dirac
semimetal is its density of states (DOS). Unlike conventional two-dimensional
massive electron systems, where the DOS is energy-independent, the DOS of a
Dirac semimetal exhibits a linear dependence on energy.

In general, as $\mathcal{B}\left(  \boldsymbol{k}\right)  $ gains an
out-of-plane component, the dispersion immediately loses topological
protection after considering the warping effect. Therefore, an in-plane Zeeman
field could tilt the Dirac cone, vanishes the Dirac point and triggers the
semimetal-insulator transition. However, under the protection of mirror
symmetry ($\phi_{B}\mathtt{=}(2n\mathtt{+}1)\pi\mathtt{/}6$ with
$n\mathtt{\in}\mathbb{Z}$), the dispersion could still remain gapless. In such
cases, the Dirac point locates on the mirror plane with $(k,\varphi
_{k})\mathtt{=}(\omega_{TI}\mathtt{/}2v_{f},\phi_{B}\mathtt{-}\pi\mathtt{/}%
2)$. Otherwise, one could follow the procedure of convex optimization to
estimate the band gap exactly.

\begin{figure*}[ptb]
\centering
\includegraphics[width=1.\linewidth]{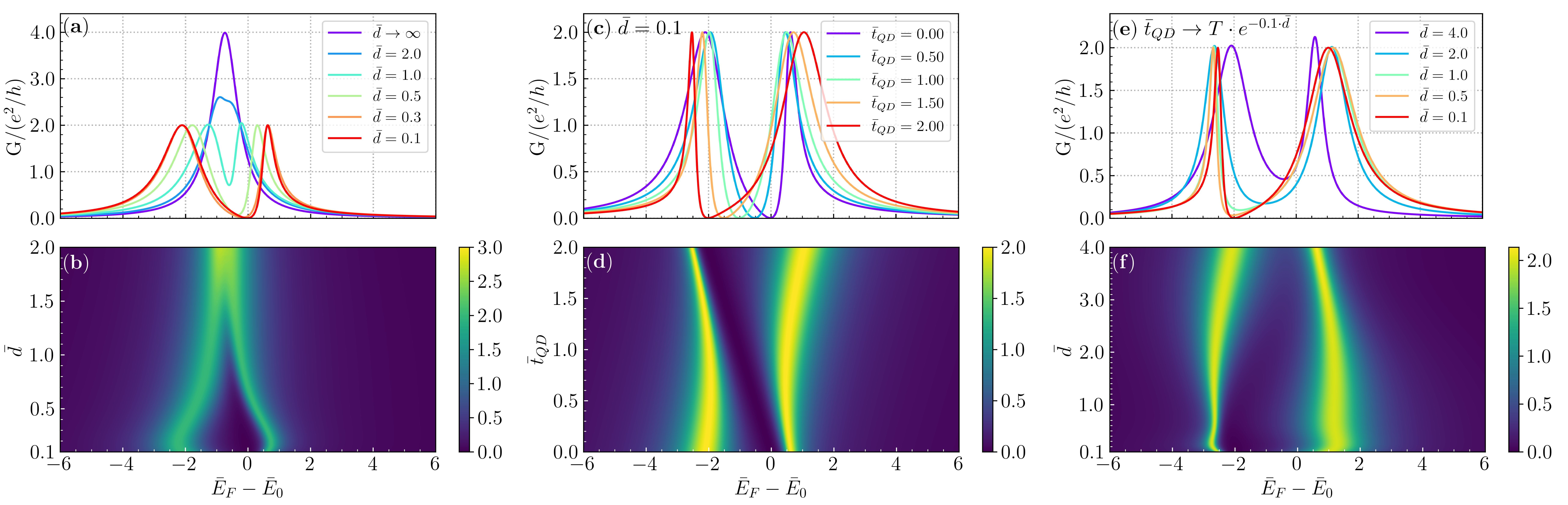} \caption{(Color online) (a) The
zero-field conductance lineshapes and (b) the corresponding trajectory as a
function of inter-dot separation $\bar{d}$. (c) The lineshape and (d) the
trajectory of the conductance as a function of inter-dot hopping strength
$\bar{t}_{QD}$ with $\bar{d}\mathtt{=}0.1$. (e) The lineshape and (f) the
trajectory of the conductance as a function of inter-dot separation $\bar{d}$
after considering $\bar{t}_{QD}\rightarrow T \cdot e^{- 0.1 \cdot\bar{d}}$
with $T\mathtt{=}2$.}%
\label{fig2}%
\end{figure*}

The conductance lineshape, governed by the Landauer-B\"{u}ttiker formula
\cite{meir1992landauer,haug2008quantum,datta1997electronic,Bruus&Flensburg}
$G\mathtt{=}T\left(  E_{F}\right)  \mathtt{\cdot}e^{2}\mathtt{/}h$, reflects
both the level spacing and the degeneracy of DQD. When a $n\mathtt{-}$fold
degenerate DQD state is aligned to the electrode Fermi energy (by adjusting
the DQD resonant energy $E_{0}$, or gate-voltage experimentally), a resonant
peak of height $n\mathtt{\cdot}e^{2}\mathtt{/}h$ emerges in conductance
lineshape. Here,
\begin{equation}
T\left(  \omega\right)  =\operatorname{Tr}\left[  \hat{\Gamma}_{L}\left(
\omega\right)  \mathcal{G}^{r}\left(  \omega\right)  \hat{\Gamma}_{R}\left(
\omega\right)  \mathcal{G}^{a}\left(  \omega\right)  \right]
\label{Transmission}%
\end{equation}
is the transmission probability and
\begin{equation}
\mathcal{G}^{r,a}\left(  \omega\right)  =\left[  \omega-\left(  H_{DQD}%
+\hat{\Sigma}^{r,a}\left(  \omega\right)  \right)  \right]  ^{-1}%
\end{equation}
denotes the DQD retarded/advanced Green's functions with
\begin{equation}
\hat{\Sigma}^{r,a}\left(  \omega\right)  =\sum\limits_{\alpha=L,R}\hat{\Sigma
}_{\alpha}^{r,a}\left(  \omega\right)  =\sum\limits_{\alpha=L,R}\hat{\Lambda
}_{\alpha}\left(  \omega\right)  \mp\frac{i}{2}\hat{\Gamma}_{\alpha}\left(
\omega\right)  \label{Sigma_analy}%
\end{equation}
the corresponding retarded/advanced self-energy. Here $\hat{\Lambda}_{\alpha}$
and $\hat{\Gamma}_{\alpha}$ are the energy-shift and decay matrices at
$\alpha$ side. (seeing Section III.A of SI for details.) By ignoring the electrodes warping effect, the regularized
self-energy could be analytically obtained with
\begin{align}
\hat{\Lambda}_{\alpha}\left(  E_{F}\right)   &  =\frac{\Gamma_{0}}{4}\tau
^{1}\otimes\left[  Y_{0}\left(  \bar{d}\right)  \sigma^{2}+iY_{1}\left(
\bar{d}\right)  \sigma^{2}\right]  ,\nonumber\\
\hat{\Gamma}_{\alpha}\left(  E_{F}\right)   &  =\frac{\Gamma_{0}}{2}\left\{
\tau^{0}\otimes\sigma^{0}+\tau^{1}\otimes\left[  J_{0}\left(  \bar{d}\right)
\sigma^{0}+iJ_{1}\left(  \bar{d}\right)  \sigma^{2}\right]  \right\}  .
\end{align}
Here, $J_{n}\left(  \bar{d}\right)  $ and $Y_{n}\left(  \bar{d}\right)  $ are
the Bessel functions of the first and the second kind, with $\bar{d}$ being
the reduced inter-dot separation and $\Gamma_{0}\mathtt{=}t^{2}\mathtt{\cdot
}k_{F}/v_{F}$ [see Fig.\thinspace\ref{fig1}(b)]. Notably, the conductance can
be expressed as
\begin{equation}
G=\frac{e^{2}}{h}\sum_{\pm}\frac{\Delta_{\pm}^{2}+\kappa\Gamma_{\pm}}%
{\Delta_{\pm}^{2}+\Gamma_{\pm}^{2}} \label{G_analy}%
\end{equation}
when the system's double point group (PG) is larger than $^{d}$C$_{1}$
\cite{sandvik2010computational}, seeing Section III.B of SI for details. In the coherent regime ($\bar{d}\mathtt{<}1$),
$\kappa\mathtt{=-}1\mathtt{+}J_{0}^{2}\left(  \bar{d}\right)  \mathtt{+}%
J_{1}^{2}\left(  \bar{d}\right)  $ rapidly decays to zero with the decreases
of $\bar{d}$. In such cases, the positions of resonance peaks and
anti-resonance dip could be located respectively, by solving $\Gamma_{\pm
}\mathtt{=}0$ and $\Delta_{\pm}\mathtt{=}0$. It is noteworthy that
Eq.\thinspace(\ref{G_analy}) encompasses most conditions except that with
extremely small inter-dot separation, because neglecting the warping term -or
equivalently, expanding the TSS effective Hamiltonian only to linear order,
-is only sufficient to describe low-energy/large-scale physics, hence failing
to capture the TSS-mediated interaction in extremely small inter-dot separations.

\emph{Results and Discussions.}---As formulated in Fig.\thinspace
\ref{fig1}(b), we start by introducing the parameters and dimensionless
variables. For TSS electrodes, the Fermi velocity $v_{F}\mathtt{=}%
255\,$meV$\mathtt{\cdot}$nm the warping factor $\lambda\mathtt{=}%
250$\thinspace meV$\mathtt{\cdot}$nm$^{3}$ are adopted \cite{Bi2Te3_1}, while
the Zeeman field intensity is scaled by $\bar{\omega}_{TI}\mathtt{=}%
\omega_{TI}/E_{F}$, with $E_{F}\mathtt{=}35\,\text{meV}$ as the Fermi energy
(for most cases, except when uncovering the universal scaling relation). For
the DQD, the inter-dot separation is measured as $\bar{d}\mathtt{=}%
d\mathtt{\cdot}k_{F}$, where $k_{F}$ is the Fermi wavevector without
considering the warping effect, given by $k_{F}\mathtt{=}E_{F}/v_{F}$. Because
we set $t\mathtt{=}1$\thinspace meV$\mathtt{\cdot}$nm for simplification,
$\Gamma_{0}$ becomes the characteristic energy scale of the TSS-mediated
interaction. Therefore, we measure the detuning between the electrode Fermi
energy and the DQD resonant energy $E_{\Delta}\mathtt{\equiv}E_{F}%
\mathtt{-}E_{0}$, the inter-dot hopping strength $t_{QD}$ and the Zeeman field
intensity $\bar{\omega}_{QD}$ by $\Gamma_{0}$, namely $\bar{E}_{\Delta
}\mathtt{=}\bar{E}_{F}\mathtt{\mathtt{-}}\bar{E}_{0}\mathtt{=}\left(
E_{F}\mathtt{-}E_{0}\right)  /\Gamma_{0}$, $\bar{t}_{QD}\mathtt{=}%
t_{QD}/\Gamma_{0}$, and $\bar{\omega}_{QD}\mathtt{=}\omega_{QD}/\Gamma_{0}$.

Akin to the 2DEG-system \cite{shahbazyan1994,shahbazyan2015}, the
zero-(magnetic) field conductance described by Eq.\thinspace(\ref{G_analy})
with
\begin{align}
\Delta_{+}  &  =\Delta_{-}=2\bar{E}_{\Delta}+J_{0}\left(  \bar{d}\right)
\left[  Y_{0}\left(  \bar{d}\right)  +2\bar{t}_{QD}\right]  +J_{1}\left(
\bar{d}\right)  Y_{1}\left(  \bar{d}\right)  ,\nonumber\\
\Gamma_{+}  &  =\Gamma_{-}=\frac{1}{2}\left\{  4\bar{E}_{\Delta}^{2}-\left[
Y_{0}\left(  \bar{d}\right)  +2\bar{t}_{QD}\right]  ^{2}-Y_{1}^{2}\left(
\bar{d}\right)  +\kappa\right\}
\end{align}
remains a double-peak lineshape of height $\mathtt{\sim}2e^{2}/h$ in the
coherent regime as shown in Fig.\thinspace\ref{fig2}(a). The two-fold
degeneracy of DQD is protected by the $^{d}$C$_{2v}$ double PG symmetry (or
$^{d}$C$_{1h}$ double PG and time-reversal symmetry when $\lambda\mathtt{\neq
}0$, see Section II of the SI for details). Beside a resonance peak of
Breit-Wigner lineshape, there is a Fano lineshape which could be traced back
to the complete set of commuting observables (CSCO). In the 2DEG-system, the
spatial and spin degrees of freedom (DoFs) are decoupled, making the system's
CSCO two-dimensional: one operator acts on the spatial DoF while the other one
acts on the spin DoF. Therefore, the self-energy matrix, along with the DQD
Green's function could be diagonalized, resulting in a conductance composed of
the superposition of two spin-degenerate resonance peaks. In contrast, in the
TSS-system, the CSCO is one-dimensional since the spatial and spin DoFs are
coupled by electrode SOC. Consequently, interference between transmission
amplitudes is allowed by the symmetry, as both the self-energy and the Green's
function could only be block-diagonalized. In such cases, a resonance of
transmission amplitude, accompanied by a $\pi$-phase shift, could introduce a
destructive interference between transmission amplitudes
\cite{joe2006classical}, and contribute to a Fano-type conductance lineshape.

On the other hand, since the symmetry remains unchanged, the conductance
retains its double-peak structure when the inter-dot hopping $\bar{t}_{QD}$ is
introduced. However, because the inter-dot hopping respects spin as a good
quantum number, while the TSS-mediated interaction does not, the interference
effect on the conductance lineshape depends on which interaction dominates. As
the hopping strength $\bar{t}_{QD}$ increases (the energy barrier between the
two dots decreases), while keeping $\bar{d}$ fixed, the Fano
effect---characterized by the asymmetry in the lineshape---becomes more
pronounced [Fig.\thinspace\ref{fig2}(c)]. Meanwhile, because the inter-dot
hopping respects the inversion symmetry and leads to DQD superposition states
with specific parity, i.e. bonding/anti-bonding states
\cite{shahbazyan2015,Zhu1,Zhu2}, while the TSS-mediated interaction breaks the
inversion symmetry, an avoided crossing trajectory is developed with the
increase of $\bar{t}_{QD}$ [Fig.\thinspace\ref{fig2}(d)].

To fully uncover the mechanism of inter-dot hopping, its distance dependence
must be taken into account. Inspired by the WKB solution of the double
potential well, we model the inter-dot hopping as $\bar{t}_{QD}%
\mathtt{\rightarrow}T\mathtt{\cdot\exp}\left(  \mathtt{-}k_{T}d\right)
\mathtt{=}T\mathtt{\cdot\exp}\left(  \mathtt{-}\bar{d}\mathtt{\cdot}%
k_{T}/k_{F}\right)  $, where $k_{T}^{-1}$ represents the characteristic
length-scale of the inter-dot hopping. Consequently, the discussion above
could be categorized into the case where $k_{F}\mathtt{\gg}k_{T}$, which
corresponds to the limit of the strong inter-dot potential barrier. We then
consider the opposite case, namely the weak barrier limit with $k_{T}/k_{F}$
$\mathtt{=}0.1$ [Fig.\thinspace\ref{fig2}(e), (f)]. Under these conditions,
the conductance lineshape could be divided into three regimes. First, the
single-peak case with $\bar{d}\mathtt{\gg}1$ [Fig.\thinspace\ref{fig2}(a)],
where all inter-dot interactions vanish. Second, the regime with $k_{F}%
^{-1}\mathtt{<}\bar{d}\mathtt{<}k_{T}^{-1}$, where the direct inter-dot
hopping dominates. In this regime, the conductance lineshape is formed by the
superposition of two resonant peaks, as the inter-dot interaction mainly
induces an energy-shift to DQD states. There is small Dicke effect with small
difference in the linewidths of the resonant states. Third, the deep
Dicke/Fano regime where $\bar{d}\mathtt{<}k_{F}^{-1}$. In this case, duo to
the electrode SOC, interference is introduced by electrode-mediated
interaction, causing the conductance lineshape to exhibit an asymmetric
lineshape with an anti-resonant dip and large difference of the linewidths of
the resonant states.\bigskip\begin{figure}[ptb]
\centering
\includegraphics[width=1.\linewidth]{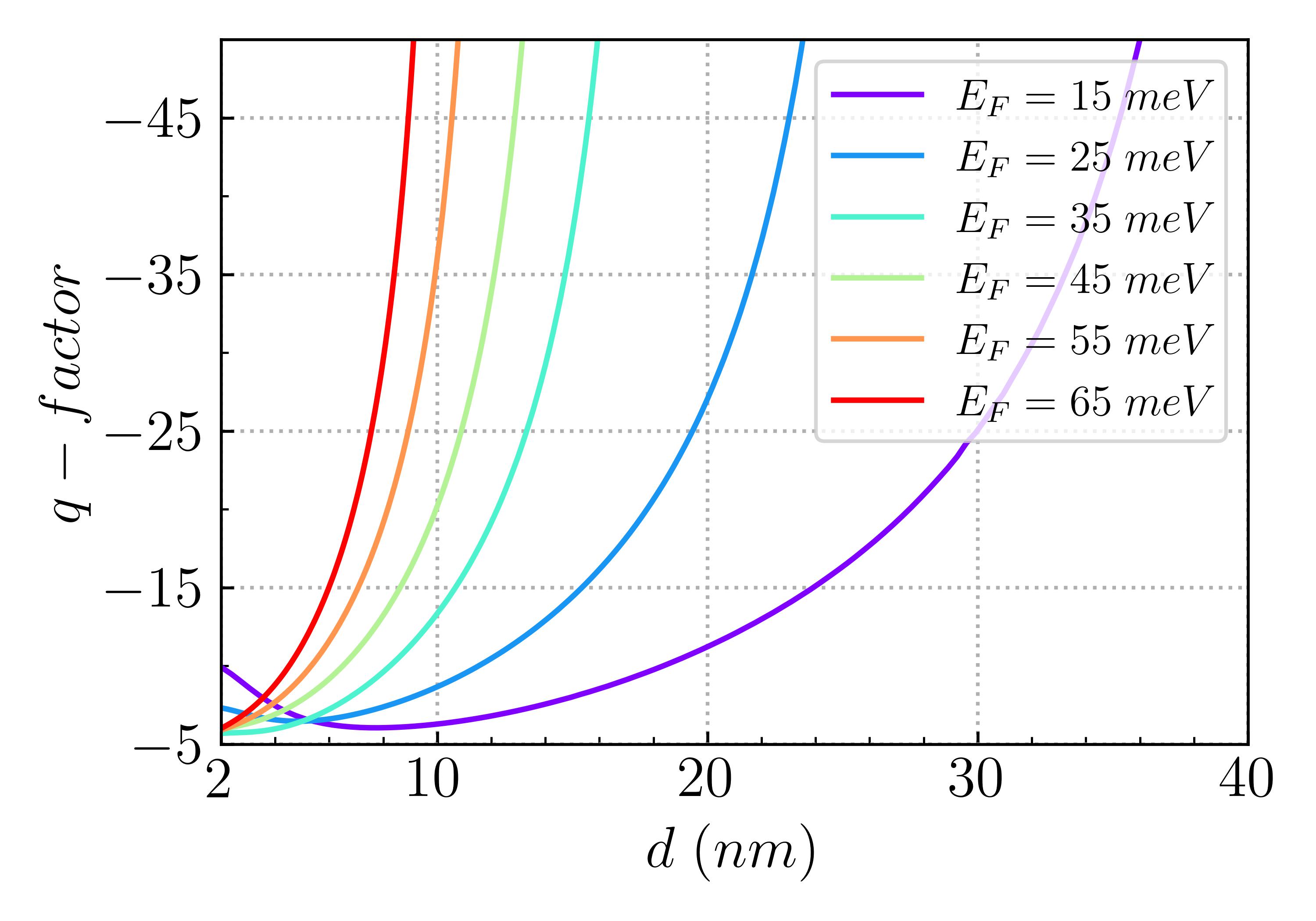} \caption{(Color online) The
Nano-ruler based on the relation between Fano factor and inter-dot distance
(for different Fermi energies).}%
\label{fig3}%
\end{figure}

\begin{figure}[ptb]
\centering
\includegraphics[width=1.\linewidth]{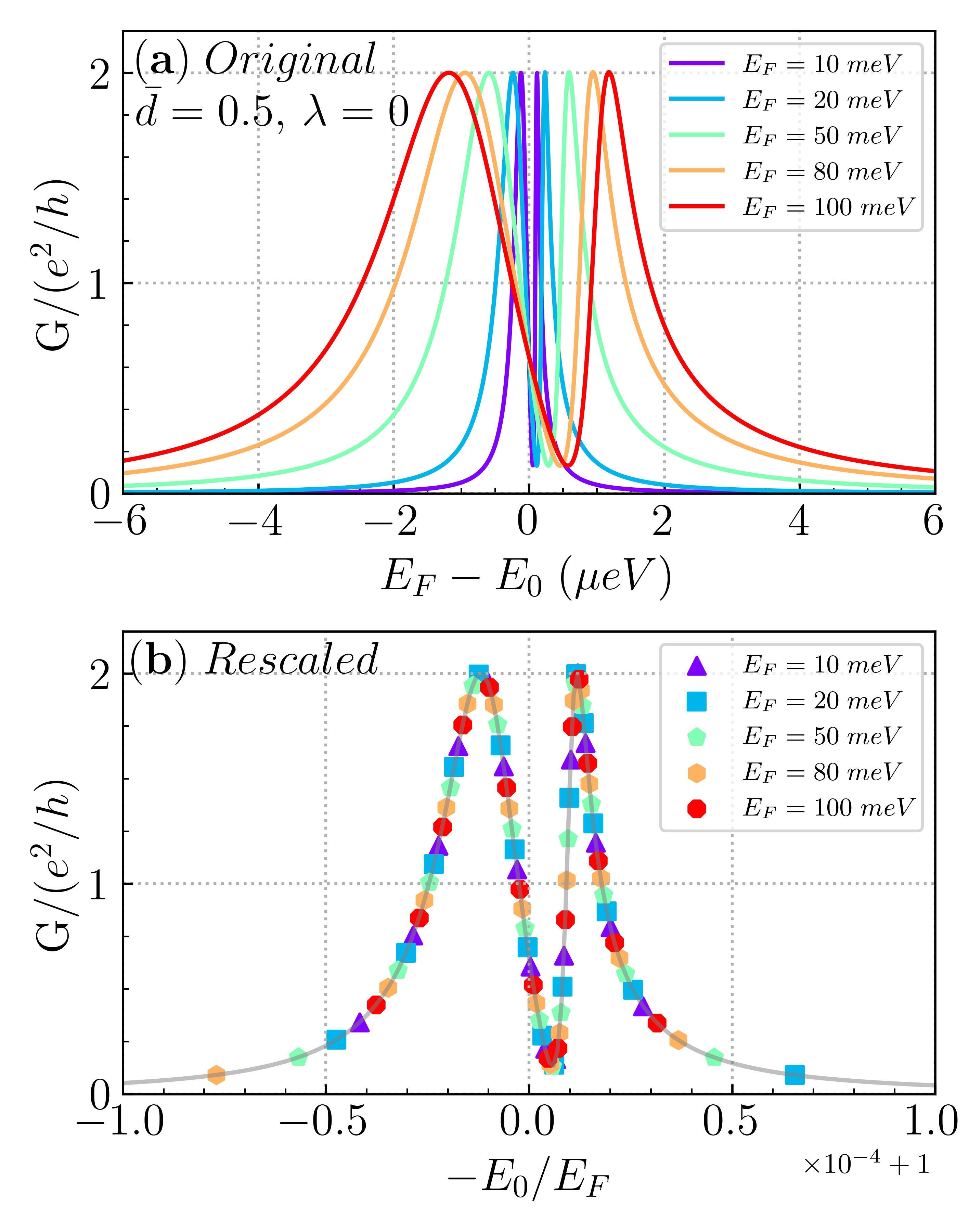} \caption{(Color online) (a) The
conductance versus $E_{0}$ for different Fermi energy $E_{F}$. (b) The
universal dependence of the conductance on the rescaled $E_{0}$ $G(-E_{0}%
/E_{F})$.}%
\label{fig4}%
\end{figure}

It is noteworthy that the asymmetric peak, along with the anti-resonant dip
could be described by the formulae of traditional Fano resonance (see Section
III.C of SI for details) \cite{FANO_RMP}. The asymmetric lineshape is
determined by the Fano $q$-factor, which depends on the inter-dot distance
through the inter-dot hopping strength and electrode-mediated interaction.
Therefore, we may develop a "nano-ruler" measuring the inter-dot distance (at
several to tens nanometers) by the conductance profile with the help of the
distance dependent Fano $q$-factor. We may measure the nanoscale distance with
high resolution/small range or lower resolution/large range with different
Fermi energy (tuned by gate voltage) as shown in Fig.\thinspace\ref{fig3}. In
applications to devices compressed or stretched, the separation of DQD, alone
with the lattice structure of electrodes, are influenced. However, due to the
protection of band topology, the electrode dispersion remains unharmed.
Therefore, by measuring the conductance and fitting the Fano $q$-factor (in
the intervals where the $q$-factor monotonically increases with $\bar{d}$),
information about DQD separation variation could be obtained. The developed
nano ruler depends on the conductance profile, thus is robust and more
accurate. Moreover, the measurement range and accuracy and be tuned by using
different gate voltage/Fermi energy.

\begin{figure*}[ptb]
\includegraphics[width=1.\linewidth]{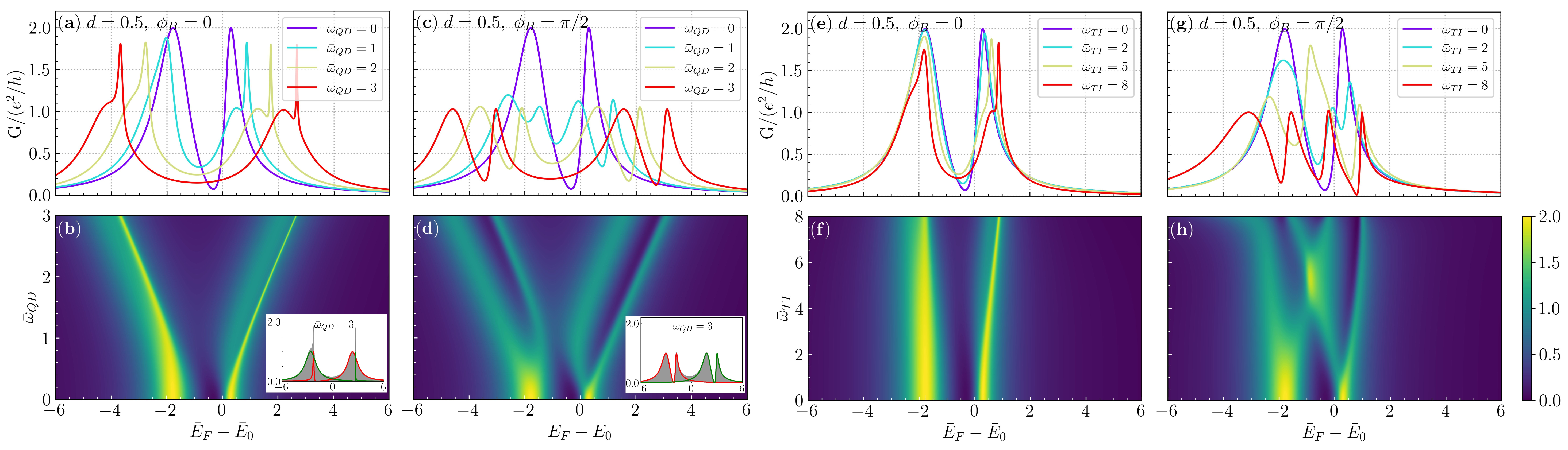} \centering
\caption{(Color online) The conductance lineshape with a vertical (a), (b) and
horizontal (c), (d) Zeeman field applied to the DQD, while the insets
illustrate the analytical results of $\bar{\omega}_{QD}\mathtt{=}3$. (e)-(h)
The corresponding cases with Zeeman fields applied to both TSS electrodes. }%
\label{fig5}%
\end{figure*}

Here we would like to emphasize that the topological nature of TSS and
associated linear dispersion results in important physical consequence.
Different from 2DEG with constant DOS, the linear energy dependence of the TSS
electrode's DOS leads to a linear dependence of the conductance linewidth on
the electrode's Fermi energy [Fig.\thinspace\ref{fig4}(a)]. Consequently,
tuning the electrode's Fermi energy (by adjusting gate-voltage) allows not
only for the adjustment of the conductance linewidth but also for determining
the peaks and dip of the conductance curve. Moreover, there is a scale
invariance. Although $G(E_{0})$ looks quite different for various $E_{F}$, the
conductance $G(E_{0}/E_{F})$ have the same lineshape for different Fermi
energy $E_{F}$ [Fig.\thinspace\ref{fig4}(b)] as a consequence of the linear
energy dependence of the DOS. It is noteworthy that the scaling relation
is a consequence of linear dispersion (associated with the TSS) and the connection
to the Fermi energy $E_{F}$. Therefore, an
interaction that introduces an $E_{F}$-independent energy scale will break the
scaling relation.

Let us now investigate the impact of an in-plane Zeeman field on the DQD.
Because the two-fold degenerate DQD\ states lose symmetry protection [PG is
lowered to $^{d}$C$_{1h}$ ($\phi_{B}\mathtt{=}n\pi\mathtt{/}2$ with
$n\mathtt{\in}\mathbb{Z}$) or $^{d}$C$_{1}$ (otherwise)], the conductance
evolves from double-peak to quad-peak lineshape with the increase of field
intensity. The lower and higher double peaks correspond to DQD states with
spin parallel and anti-parallel to the Zeeman field, respectively. Aside from
influencing the spin orientation of the DQD states, the Zeeman field also
regulates the interference. For the case with vertical field ($\phi
_{B}\mathtt{=}0$) [Fig.\thinspace\ref{fig5}(a)], the Zeeman field leads to
both the shift of the peaks and modulation of the peak width due to the TSS
mediated interaction as shown by Eq.\thinspace(\ref{G_analy}) with%
\begin{align}
\Delta_{\pm}  &  =2\bar{E}_{\Delta}+J_{0}\left(  \bar{d}\right)  \left[
Y_{0}\left(  \bar{d}\right)  \pm2\bar{\omega}_{QD}\right]  +J_{1}\left(
\bar{d}\right)  Y_{1}\left(  \bar{d}\right)  ,\nonumber\\
\Gamma_{\pm}  &  =\frac{1}{2}\left[  4\bar{E}_{\Delta}^{2}-\left[
Y_{0}\left(  \bar{d}\right)  \pm2\bar{\omega}_{QD}\right]  ^{2}-Y_{1}%
^{2}\left(  \bar{d}\right)  +\kappa\right]  .
\end{align}
While in the horizontal ($\phi_{B}\mathtt{=}\pi\mathtt{/}2$) field case, as
shown in Fig.\thinspace\ref{fig5}(c), the conductance consists of a pair of
Fano-type double-peaks described by Eq.\thinspace(\ref{G_analy}) with%
\begin{align}
\Delta_{\pm}  &  =2\left(  \bar{E}_{\Delta}\pm\bar{\omega}_{QD}\right)
+J_{0}\left(  \bar{d}\right)  Y_{0}\left(  \bar{d}\right)  +J_{1}\left(
\bar{d}\right)  Y_{1}\left(  \bar{d}\right)  ,\nonumber\\
\Gamma_{\pm}  &  =\frac{1}{2}\left[  4\left(  \bar{E}_{\Delta}\pm\bar{\omega
}_{QD}\right)  ^{2}-Y_{0}^{2}\left(  \bar{d}\right)  -Y_{1}^{2}\left(  \bar
{d}\right)  +\kappa\right]  .
\end{align}
This result suggests that interference takes place between DQD states with the
same spin. The Zeeman field mainly leads to the splitting/shift of the peaks.
By separately plotting each of the Fano-type double-peaks that make up the
quad-peak lineshape [insets of Figs.\thinspace\ref{fig5}(b) and \ref{fig5}%
(d)], we find that the interference between DQD states with opposite and same
spins appears in the case with $\phi_{B}\mathtt{=}0$ and $\phi_{B}%
\mathtt{=}\pi/2$, respectively. Such outcomes are consistent with the results
of block-diagonalization.

\begin{figure}[ptb]
\centering
\includegraphics[width=1.\linewidth]{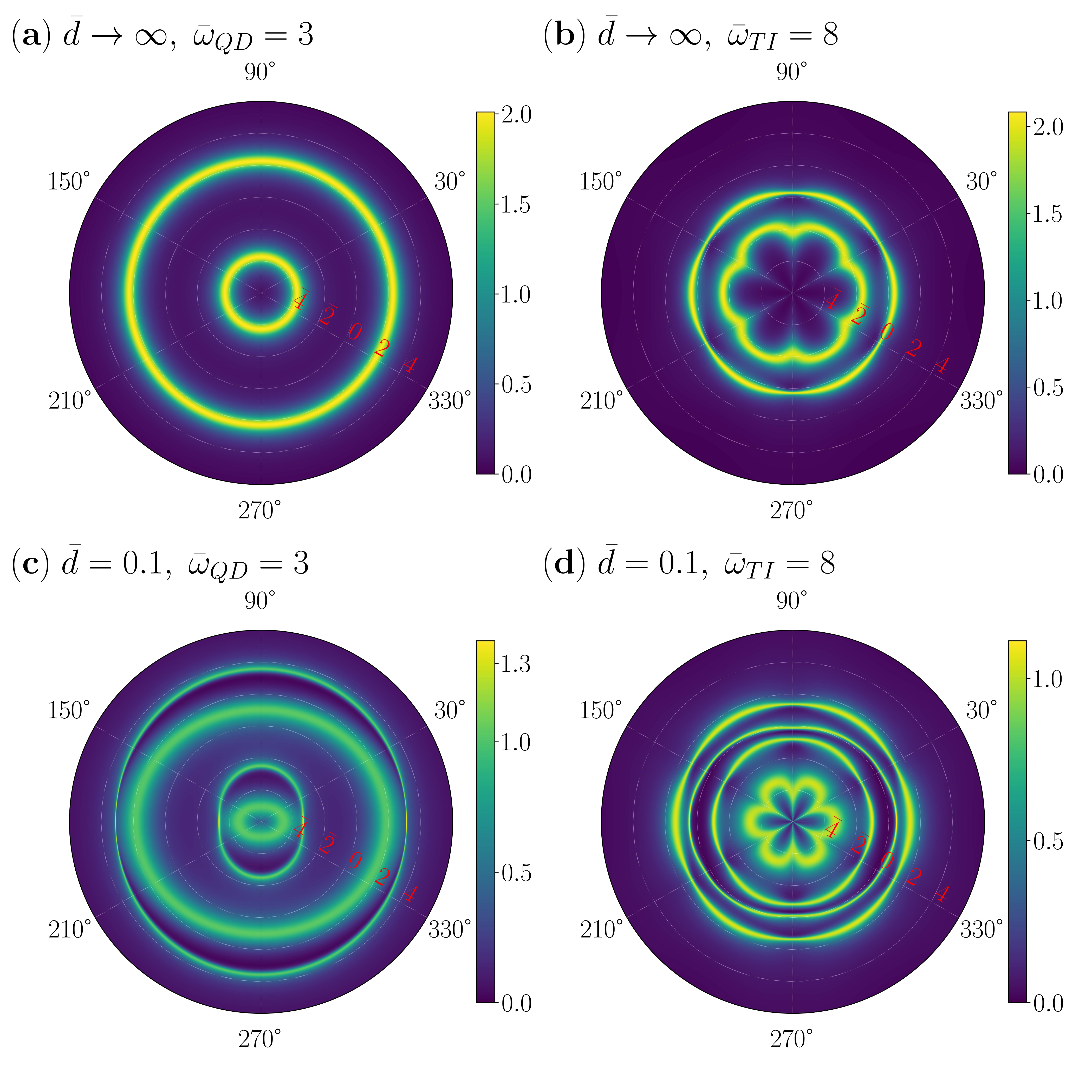} \caption{ (Color online) The
conductance trajectory as a function of field orientation in the single QD
cases ($\bar{d}\mathtt{\rightarrow}\infty$) with the Zeeman field applied to
DQD (a) and both electrodes (b). (c), (d) The corresponding cases with finite
inter-dot separation. The polar axis and angle are labeled by $\bar{E}%
_{F}\mathtt{-}\bar{E}_{0}$ and $\phi_{B}$, respectively (The negative sign is
written above the number).}%
\label{fig6}%
\end{figure}

Furthermore, we explore the impact of the in-plane Zeeman fields on the
distortion of the TSS dispersion of the electrodes, also known as the tilting
of the Dirac cone \cite{TiltDirac1,TiltDirac2,TiltDirac3}. The corresponding
conductance results are shown in Figs.\thinspace\ref{fig5}(e)-\ref{fig5}(h).
The Zeeman field influences the electrode dispersion by collaborating with the
hexagonal warping effect (a high-order effect $\mathtt{\sim}k^{3}$), resulting
in a less efficient spin-splitting of the DQD levels. Applying Zeeman fields
to both electrodes directly impacts the DQD-electrode interaction $H_{int}$,
leading to a more efficient modulation of interference. For instance, in the
single QD case (i.e., $\bar{d}\mathtt{\gg}1$), an asymmetric Fano-type
lineshape could be achieved by applying Zeeman fields to both electrodes
instead of just to the DQD, see Fig.\thinspace\ref{fig6}(b).

When the Zeeman field is applied to the DQD, it competes with the TSS-mediated
interaction because it protects the $U(1)$ spin rotational symmetry along the
field, while the latter does not. Taking into account the three-fold
rotational symmetry, the conductance shows interesting spatial dependence on
the magnetic field orientation $\phi_{B}$. Apart from the intrinsic
$\pi\mathtt{-}$periodicity stemming from TR symmetry, the periodicity of
conductance reflects the rotational symmetry of the system. In the content of
a single QD where the system possesses $^{d}$C$_{3v}$ PG symmetry, as shown in
Fig.\thinspace\ref{fig6}(b), the conductance demonstrates a hexagonal pattern
when the Zeeman fields are equally applied to both electrodes. The periodicity
can be decomposed as
\begin{equation}
G\left(  \phi_{B}\right)  =G\left(  \phi_{B}+n_{1}\pi+\frac{2}{3}n_{2}%
\pi\right)  , \label{period}%
\end{equation}
where $n_{1/2}\mathtt{\in}\mathbb{Z}$. One could easily confirm that it
renders a periodicity of $\pi/3$, and the detailed derivations of Eq.
(\ref{period}) are shown in Section IV of the SI. However, when the Zeeman
field is applied to the DQD, the conductance only displays a circular
trajectory [Fig.\thinspace\ref{fig6}(a)], indicating that the influence of
electrode warping effect is negligible unless strong Zeeman fields are
directly applied to both electrodes. On the other hand, because the system
possesses, at most, two-fold rotational symmetry in the cases of finite
inter-dot separation, the conductance always displays a two-fold rotational
symmetric trajectory, regardless of where the Zeeman field is applied, as
shown in Figs.\thinspace\ref{fig6}(c) and \ref{fig6}(d).

\emph{Conclusions.}---In this work, we studied the resonant tunneling of a
parallel DQD sandwiched between TSS. Without a magnetic field, the conductance
exhibits Dicke effect and a Fano-type asymmetrical lineshape. Furthermore,
nano-rules with different measurement range and resolution based on the Fano
$q$-factor have been developed. A universal scaling relation of the
conductance has been found. When an in-plane Zeeman field is applied, the
conductance changes from a double-peak shape to a quad-peak shape and the
dependence of the conductance on the orientation of the magnetic field reveals
the rotational symmetry of the system. Our findings deepen the understanding
of quantum transport in the presence of SOC and topological states, with
potential applications in quantum information etc..

We would like to thank Y.-M. Zhao, W. Chen, Z.-H. Zhou and X. Wang for helpful
discussions. This work was supported by the National Natural Science
Foundation of China (12175023, 12174032, 12275031). National Natural Science
Foundation of China-Research Grants Council (11861161002); National Key
Research and Development Program of China (2017YFA0303400).


\begin{thebibliography}{45}
	\expandafter\ifx\csname natexlab\endcsname\relax\def\natexlab#1{#1}\fi
	\expandafter\ifx\csname bibnamefont\endcsname\relax
	\def\bibnamefont#1{#1}\fi
	\expandafter\ifx\csname bibfnamefont\endcsname\relax
	\def\bibfnamefont#1{#1}\fi
	\expandafter\ifx\csname citenamefont\endcsname\relax
	\def\citenamefont#1{#1}\fi
	\expandafter\ifx\csname url\endcsname\relax
	\def\url#1{\texttt{#1}}\fi
	\expandafter\ifx\csname urlprefix\endcsname\relax\def\urlprefix{URL }\fi
	\providecommand{\bibinfo}[2]{#2}
	\providecommand{\eprint}[2][]{\url{#2}}
	
	\bibitem[{\citenamefont{Hanson et~al.}(2007)\citenamefont{Hanson, Kouwenhoven,
			Petta, Tarucha, and Vandersypen}}]{QD_RMP}
	\bibinfo{author}{\bibfnamefont{R.}~\bibnamefont{Hanson}},
	\bibinfo{author}{\bibfnamefont{L.~P.} \bibnamefont{Kouwenhoven}},
	\bibinfo{author}{\bibfnamefont{J.~R.} \bibnamefont{Petta}},
	\bibinfo{author}{\bibfnamefont{S.}~\bibnamefont{Tarucha}}, \bibnamefont{and}
	\bibinfo{author}{\bibfnamefont{L.~M.} \bibnamefont{Vandersypen}},
	\bibinfo{journal}{Rev. Mod. Phys.} \textbf{\bibinfo{volume}{79}},
	\bibinfo{pages}{1217} (\bibinfo{year}{2007}).
	
	\bibitem[{\citenamefont{Zhang et~al.}(2003)\citenamefont{Zhang, Xue, and
			Xie}}]{SpinQD1}
	\bibinfo{author}{\bibfnamefont{P.}~\bibnamefont{Zhang}},
	\bibinfo{author}{\bibfnamefont{Q.-K.} \bibnamefont{Xue}}, \bibnamefont{and}
	\bibinfo{author}{\bibfnamefont{X.}~\bibnamefont{Xie}},
	\bibinfo{journal}{Phys. Rev. Lett.} \textbf{\bibinfo{volume}{91}},
	\bibinfo{pages}{196602} (\bibinfo{year}{2003}).
	
	\bibitem[{\citenamefont{Wang et~al.}(2004)\citenamefont{Wang, Sun, and
			Guo}}]{SpinQD2}
	\bibinfo{author}{\bibfnamefont{D.-K.} \bibnamefont{Wang}},
	\bibinfo{author}{\bibfnamefont{Q.-f.} \bibnamefont{Sun}}, \bibnamefont{and}
	\bibinfo{author}{\bibfnamefont{H.}~\bibnamefont{Guo}},
	\bibinfo{journal}{Phys. Rev. B} \textbf{\bibinfo{volume}{69}},
	\bibinfo{pages}{205312} (\bibinfo{year}{2004}).
	
	\bibitem[{\citenamefont{Awschalom et~al.}(2013)\citenamefont{Awschalom,
			Bassett, Dzurak, Hu, and Petta}}]{SpinQD3}
	\bibinfo{author}{\bibfnamefont{D.~D.} \bibnamefont{Awschalom}},
	\bibinfo{author}{\bibfnamefont{L.~C.} \bibnamefont{Bassett}},
	\bibinfo{author}{\bibfnamefont{A.~S.} \bibnamefont{Dzurak}},
	\bibinfo{author}{\bibfnamefont{E.~L.} \bibnamefont{Hu}}, \bibnamefont{and}
	\bibinfo{author}{\bibfnamefont{J.~R.} \bibnamefont{Petta}},
	\bibinfo{journal}{Science} \textbf{\bibinfo{volume}{339}},
	\bibinfo{pages}{1174} (\bibinfo{year}{2013}).
	
	\bibitem[{\citenamefont{Li et~al.}(2017)\citenamefont{Li, Kang, Caroff, and
			Xu}}]{QDcoherence1}
	\bibinfo{author}{\bibfnamefont{S.}~\bibnamefont{Li}},
	\bibinfo{author}{\bibfnamefont{N.}~\bibnamefont{Kang}},
	\bibinfo{author}{\bibfnamefont{P.}~\bibnamefont{Caroff}}, \bibnamefont{and}
	\bibinfo{author}{\bibfnamefont{H.}~\bibnamefont{Xu}}, \bibinfo{journal}{Phys.
		Rev. B} \textbf{\bibinfo{volume}{95}}, \bibinfo{pages}{014515}
	(\bibinfo{year}{2017}).
	
	\bibitem[{\citenamefont{Debbarma et~al.}(2022)\citenamefont{Debbarma, Aspegren,
			Bostr{\"o}m, Lehmann, Dick, and Thelander}}]{QDcoherence2}
	\bibinfo{author}{\bibfnamefont{R.}~\bibnamefont{Debbarma}},
	\bibinfo{author}{\bibfnamefont{M.}~\bibnamefont{Aspegren}},
	\bibinfo{author}{\bibfnamefont{F.~V.} \bibnamefont{Bostr{\"o}m}},
	\bibinfo{author}{\bibfnamefont{S.}~\bibnamefont{Lehmann}},
	\bibinfo{author}{\bibfnamefont{K.}~\bibnamefont{Dick}}, \bibnamefont{and}
	\bibinfo{author}{\bibfnamefont{C.}~\bibnamefont{Thelander}},
	\bibinfo{journal}{Phys. Rev. B} \textbf{\bibinfo{volume}{106}},
	\bibinfo{pages}{L180507} (\bibinfo{year}{2022}).
	
	\bibitem[{\citenamefont{Thomas et~al.}(2021)\citenamefont{Thomas, Nilsson,
			Ciaccia, J{\"u}nger, Rossi, Zannier, Sorba, Baumgartner, and
			Sch{\"o}nenberger}}]{QDcoherence3}
	\bibinfo{author}{\bibfnamefont{F.~S.} \bibnamefont{Thomas}},
	\bibinfo{author}{\bibfnamefont{M.}~\bibnamefont{Nilsson}},
	\bibinfo{author}{\bibfnamefont{C.}~\bibnamefont{Ciaccia}},
	\bibinfo{author}{\bibfnamefont{C.}~\bibnamefont{J{\"u}nger}},
	\bibinfo{author}{\bibfnamefont{F.}~\bibnamefont{Rossi}},
	\bibinfo{author}{\bibfnamefont{V.}~\bibnamefont{Zannier}},
	\bibinfo{author}{\bibfnamefont{L.}~\bibnamefont{Sorba}},
	\bibinfo{author}{\bibfnamefont{A.}~\bibnamefont{Baumgartner}},
	\bibnamefont{and}
	\bibinfo{author}{\bibfnamefont{C.}~\bibnamefont{Sch{\"o}nenberger}},
	\bibinfo{journal}{Phys. Rev. B} \textbf{\bibinfo{volume}{104}},
	\bibinfo{pages}{115415} (\bibinfo{year}{2021}).
	
	\bibitem[{\citenamefont{Wang et~al.}(2024)\citenamefont{Wang, Wan, Teale,
			Grater, Zhao, Zhang, Duan, Imran, Wang, Hoogland et~al.}}]{WangYK2024}
	\bibinfo{author}{\bibfnamefont{Y.-K.} \bibnamefont{Wang}},
	\bibinfo{author}{\bibfnamefont{H.}~\bibnamefont{Wan}},
	\bibinfo{author}{\bibfnamefont{S.}~\bibnamefont{Teale}},
	\bibinfo{author}{\bibfnamefont{L.}~\bibnamefont{Grater}},
	\bibinfo{author}{\bibfnamefont{F.}~\bibnamefont{Zhao}},
	\bibinfo{author}{\bibfnamefont{Z.}~\bibnamefont{Zhang}},
	\bibinfo{author}{\bibfnamefont{H.-W.} \bibnamefont{Duan}},
	\bibinfo{author}{\bibfnamefont{M.}~\bibnamefont{Imran}},
	\bibinfo{author}{\bibfnamefont{S.-D.} \bibnamefont{Wang}},
	\bibinfo{author}{\bibfnamefont{S.}~\bibnamefont{Hoogland}},
	\bibnamefont{et~al.}, \bibinfo{journal}{Nature}
	\textbf{\bibinfo{volume}{629}}, \bibinfo{pages}{586} (\bibinfo{year}{2024}).
	
	\bibitem[{\citenamefont{Lu et~al.}(2005)\citenamefont{Lu, L{\"u}, and
			Zhu}}]{Zhu1}
	\bibinfo{author}{\bibfnamefont{H.}~\bibnamefont{Lu}},
	\bibinfo{author}{\bibfnamefont{R.}~\bibnamefont{L{\"u}}}, \bibnamefont{and}
	\bibinfo{author}{\bibfnamefont{B.-F.} \bibnamefont{Zhu}},
	\bibinfo{journal}{Phys. Rev. B} \textbf{\bibinfo{volume}{71}},
	\bibinfo{pages}{235320} (\bibinfo{year}{2005}).
	
	\bibitem[{\citenamefont{Lu et~al.}(2006)\citenamefont{Lu, Lu, and Zhu}}]{Zhu2}
	\bibinfo{author}{\bibfnamefont{H.}~\bibnamefont{Lu}},
	\bibinfo{author}{\bibfnamefont{R.}~\bibnamefont{Lu}}, \bibnamefont{and}
	\bibinfo{author}{\bibfnamefont{B.-F.} \bibnamefont{Zhu}},
	\bibinfo{journal}{J. Phys. Condens. Matter} \textbf{\bibinfo{volume}{18}},
	\bibinfo{pages}{8961} (\bibinfo{year}{2006}).
	
	\bibitem[{\citenamefont{Ivanov}(2017)}]{MZMQD1}
	\bibinfo{author}{\bibfnamefont{T.}~\bibnamefont{Ivanov}},
	\bibinfo{journal}{Phys. Rev. B} \textbf{\bibinfo{volume}{96}},
	\bibinfo{pages}{035417} (\bibinfo{year}{2017}).
	
	\bibitem[{\citenamefont{Cifuentes and da~Silva}(2019)}]{MZMQD2}
	\bibinfo{author}{\bibfnamefont{J.~D.} \bibnamefont{Cifuentes}}
	\bibnamefont{and} \bibinfo{author}{\bibfnamefont{L.~G.~D.}
		\bibnamefont{da~Silva}}, \bibinfo{journal}{Phys. Rev. B}
	\textbf{\bibinfo{volume}{100}}, \bibinfo{pages}{085429}
	(\bibinfo{year}{2019}).
	
	\bibitem[{\citenamefont{Tsintzis et~al.}(2022)\citenamefont{Tsintzis, Souto,
			and Leijnse}}]{MZMQD3}
	\bibinfo{author}{\bibfnamefont{A.}~\bibnamefont{Tsintzis}},
	\bibinfo{author}{\bibfnamefont{R.~S.} \bibnamefont{Souto}}, \bibnamefont{and}
	\bibinfo{author}{\bibfnamefont{M.}~\bibnamefont{Leijnse}},
	\bibinfo{journal}{Phys. Rev. B} \textbf{\bibinfo{volume}{106}},
	\bibinfo{pages}{L201404} (\bibinfo{year}{2022}).
	
	\bibitem[{\citenamefont{P{\"o}schl et~al.}(2022)\citenamefont{P{\"o}schl,
			Danilenko, Sabonis, Kristjuhan, Lindemann, Thomas, Manfra, and
			Marcus}}]{MZMQD4}
	\bibinfo{author}{\bibfnamefont{A.}~\bibnamefont{P{\"o}schl}},
	\bibinfo{author}{\bibfnamefont{A.}~\bibnamefont{Danilenko}},
	\bibinfo{author}{\bibfnamefont{D.}~\bibnamefont{Sabonis}},
	\bibinfo{author}{\bibfnamefont{K.}~\bibnamefont{Kristjuhan}},
	\bibinfo{author}{\bibfnamefont{T.}~\bibnamefont{Lindemann}},
	\bibinfo{author}{\bibfnamefont{C.}~\bibnamefont{Thomas}},
	\bibinfo{author}{\bibfnamefont{M.~J.} \bibnamefont{Manfra}},
	\bibnamefont{and} \bibinfo{author}{\bibfnamefont{C.~M.}
		\bibnamefont{Marcus}}, \bibinfo{journal}{Phys. Rev. B}
	\textbf{\bibinfo{volume}{106}}, \bibinfo{pages}{L161301}
	(\bibinfo{year}{2022}).
	
	\bibitem[{\citenamefont{Matern et~al.}(2023)\citenamefont{Matern, Macieszczak,
			Wozny, and Leijnse}}]{matern2023metastability}
	\bibinfo{author}{\bibfnamefont{S.}~\bibnamefont{Matern}},
	\bibinfo{author}{\bibfnamefont{K.}~\bibnamefont{Macieszczak}},
	\bibinfo{author}{\bibfnamefont{S.}~\bibnamefont{Wozny}}, \bibnamefont{and}
	\bibinfo{author}{\bibfnamefont{M.}~\bibnamefont{Leijnse}},
	\bibinfo{journal}{Phys. Rev. B} \textbf{\bibinfo{volume}{107}},
	\bibinfo{pages}{125424} (\bibinfo{year}{2023}).
	
	\bibitem[{\citenamefont{Majek et~al.}(2022)\citenamefont{Majek, W{\'o}jcik, and
			Weymann}}]{majek2022spin}
	\bibinfo{author}{\bibfnamefont{P.}~\bibnamefont{Majek}},
	\bibinfo{author}{\bibfnamefont{K.~P.} \bibnamefont{W{\'o}jcik}},
	\bibnamefont{and} \bibinfo{author}{\bibfnamefont{I.}~\bibnamefont{Weymann}},
	\bibinfo{journal}{Phys. Rev. B} \textbf{\bibinfo{volume}{105}},
	\bibinfo{pages}{075418} (\bibinfo{year}{2022}).
	
	\bibitem[{\citenamefont{Feng et~al.}(2023)\citenamefont{Feng, Yoneda, Huang,
			Su, Tanttu, Yang, Cifuentes, Chan, Gilbert, Leon et~al.}}]{feng2023control}
	\bibinfo{author}{\bibfnamefont{M.}~\bibnamefont{Feng}},
	\bibinfo{author}{\bibfnamefont{J.}~\bibnamefont{Yoneda}},
	\bibinfo{author}{\bibfnamefont{W.}~\bibnamefont{Huang}},
	\bibinfo{author}{\bibfnamefont{Y.}~\bibnamefont{Su}},
	\bibinfo{author}{\bibfnamefont{T.}~\bibnamefont{Tanttu}},
	\bibinfo{author}{\bibfnamefont{C.~H.} \bibnamefont{Yang}},
	\bibinfo{author}{\bibfnamefont{J.~D.} \bibnamefont{Cifuentes}},
	\bibinfo{author}{\bibfnamefont{K.~W.} \bibnamefont{Chan}},
	\bibinfo{author}{\bibfnamefont{W.}~\bibnamefont{Gilbert}},
	\bibinfo{author}{\bibfnamefont{R.~C.} \bibnamefont{Leon}},
	\bibnamefont{et~al.}, \bibinfo{journal}{Phys. Rev. B}
	\textbf{\bibinfo{volume}{107}}, \bibinfo{pages}{085427}
	(\bibinfo{year}{2023}).
	
	\bibitem[{\citenamefont{Shahbazyan and Raikh}(1994)}]{shahbazyan1994}
	\bibinfo{author}{\bibfnamefont{T.}~\bibnamefont{Shahbazyan}} \bibnamefont{and}
	\bibinfo{author}{\bibfnamefont{M.}~\bibnamefont{Raikh}},
	\bibinfo{journal}{Phys. Rev. B} \textbf{\bibinfo{volume}{49}},
	\bibinfo{pages}{17123} (\bibinfo{year}{1994}).
	
	\bibitem[{\citenamefont{Petrosyan et~al.}(2011)\citenamefont{Petrosyan,
			Kirakosyan, and Shahbazyan}}]{shahbazyan2011}
	\bibinfo{author}{\bibfnamefont{L.}~\bibnamefont{Petrosyan}},
	\bibinfo{author}{\bibfnamefont{A.}~\bibnamefont{Kirakosyan}},
	\bibnamefont{and}
	\bibinfo{author}{\bibfnamefont{T.}~\bibnamefont{Shahbazyan}},
	\bibinfo{journal}{Phys. Rev. Lett.} \textbf{\bibinfo{volume}{107}},
	\bibinfo{pages}{196802} (\bibinfo{year}{2011}).
	
	\bibitem[{\citenamefont{Orellana et~al.}(2004)\citenamefont{Orellana,
			de~Guevara, and Claro}}]{Dicke_Fano1}
	\bibinfo{author}{\bibfnamefont{P.~A.} \bibnamefont{Orellana}},
	\bibinfo{author}{\bibfnamefont{M.~L.} \bibnamefont{de~Guevara}},
	\bibnamefont{and} \bibinfo{author}{\bibfnamefont{F.}~\bibnamefont{Claro}},
	\bibinfo{journal}{Physical Review B} \textbf{\bibinfo{volume}{70}},
	\bibinfo{pages}{233315} (\bibinfo{year}{2004}).
	
	\bibitem[{\citenamefont{Ojeda et~al.}(2009)\citenamefont{Ojeda, Pacheco, and
			Orellana}}]{Dicke_Fano2}
	\bibinfo{author}{\bibfnamefont{J.~H.} \bibnamefont{Ojeda}},
	\bibinfo{author}{\bibfnamefont{M.}~\bibnamefont{Pacheco}}, \bibnamefont{and}
	\bibinfo{author}{\bibfnamefont{P.~A.} \bibnamefont{Orellana}},
	\bibinfo{journal}{Nanotechnology} \textbf{\bibinfo{volume}{20}},
	\bibinfo{pages}{434013} (\bibinfo{year}{2009}).
	
	\bibitem[{\citenamefont{Gonz{\'a}lez~I
			et~al.}(2021)\citenamefont{Gonz{\'a}lez~I, Pacheco, Calle, Siqueira, and
			Orellana}}]{Dicke_Fano3}
	\bibinfo{author}{\bibfnamefont{A.}~\bibnamefont{Gonz{\'a}lez~I}},
	\bibinfo{author}{\bibfnamefont{M.}~\bibnamefont{Pacheco}},
	\bibinfo{author}{\bibfnamefont{A.}~\bibnamefont{Calle}},
	\bibinfo{author}{\bibfnamefont{E.}~\bibnamefont{Siqueira}}, \bibnamefont{and}
	\bibinfo{author}{\bibfnamefont{P.}~\bibnamefont{Orellana}},
	\bibinfo{journal}{Scientific Reports} \textbf{\bibinfo{volume}{11}},
	\bibinfo{pages}{3941} (\bibinfo{year}{2021}).
	
	\bibitem[{\citenamefont{Wang et~al.}(2022)\citenamefont{Wang, Zhou, Yan, Li,
			Nan, Zhang, Ma, Wang, Ma, Luo et~al.}}]{Dicke_Fano4}
	\bibinfo{author}{\bibfnamefont{J.-N.} \bibnamefont{Wang}},
	\bibinfo{author}{\bibfnamefont{W.-H.} \bibnamefont{Zhou}},
	\bibinfo{author}{\bibfnamefont{Y.-X.} \bibnamefont{Yan}},
	\bibinfo{author}{\bibfnamefont{W.}~\bibnamefont{Li}},
	\bibinfo{author}{\bibfnamefont{N.}~\bibnamefont{Nan}},
	\bibinfo{author}{\bibfnamefont{J.}~\bibnamefont{Zhang}},
	\bibinfo{author}{\bibfnamefont{Y.-N.} \bibnamefont{Ma}},
	\bibinfo{author}{\bibfnamefont{P.-C.} \bibnamefont{Wang}},
	\bibinfo{author}{\bibfnamefont{X.-R.} \bibnamefont{Ma}},
	\bibinfo{author}{\bibfnamefont{S.-J.} \bibnamefont{Luo}},
	\bibnamefont{et~al.}, \bibinfo{journal}{Phys. Rev. B}
	\textbf{\bibinfo{volume}{106}}, \bibinfo{pages}{035428}
	(\bibinfo{year}{2022}).
	
	\bibitem[{\citenamefont{Petrosyan and Shahbazyan}(2015)}]{shahbazyan2015}
	\bibinfo{author}{\bibfnamefont{L.}~\bibnamefont{Petrosyan}} \bibnamefont{and}
	\bibinfo{author}{\bibfnamefont{T.}~\bibnamefont{Shahbazyan}},
	\bibinfo{journal}{Phys. Rev. B} \textbf{\bibinfo{volume}{92}},
	\bibinfo{pages}{115423} (\bibinfo{year}{2015}).
	
	\bibitem[{\citenamefont{Hasan and Kane}(2010)}]{hasan}
	\bibinfo{author}{\bibfnamefont{M.~Z.} \bibnamefont{Hasan}} \bibnamefont{and}
	\bibinfo{author}{\bibfnamefont{C.~L.} \bibnamefont{Kane}},
	\bibinfo{journal}{Rev. Mod. Phys.} \textbf{\bibinfo{volume}{82}},
	\bibinfo{pages}{3045} (\bibinfo{year}{2010}).
	
	\bibitem[{\citenamefont{Qi and Zhang}(2011)}]{SCZhang}
	\bibinfo{author}{\bibfnamefont{X.-L.} \bibnamefont{Qi}} \bibnamefont{and}
	\bibinfo{author}{\bibfnamefont{S.-C.} \bibnamefont{Zhang}},
	\bibinfo{journal}{Rev. Mod. Phys.} \textbf{\bibinfo{volume}{83}},
	\bibinfo{pages}{1057} (\bibinfo{year}{2011}).
	
	\bibitem[{\citenamefont{Shen}(2012)}]{shen2012topological}
	\bibinfo{author}{\bibfnamefont{S.-Q.} \bibnamefont{Shen}},
	\emph{\bibinfo{title}{Topological insulators}}, vol. \bibinfo{volume}{174}
	(\bibinfo{publisher}{Springer}, \bibinfo{year}{2012}).
	
	\bibitem[{\citenamefont{Fu et~al.}(2007)\citenamefont{Fu, Kane, and
			Mele}}]{3DTI}
	\bibinfo{author}{\bibfnamefont{L.}~\bibnamefont{Fu}},
	\bibinfo{author}{\bibfnamefont{C.~L.} \bibnamefont{Kane}}, \bibnamefont{and}
	\bibinfo{author}{\bibfnamefont{E.~J.} \bibnamefont{Mele}},
	\bibinfo{journal}{Phys. Rev. Lett.} \textbf{\bibinfo{volume}{98}},
	\bibinfo{pages}{106803} (\bibinfo{year}{2007}).
	
	\bibitem[{\citenamefont{Liu et~al.}(2010)\citenamefont{Liu, Qi, Zhang, Dai,
			Fang, and Zhang}}]{modelTI}
	\bibinfo{author}{\bibfnamefont{C.-X.} \bibnamefont{Liu}},
	\bibinfo{author}{\bibfnamefont{X.-L.} \bibnamefont{Qi}},
	\bibinfo{author}{\bibfnamefont{H.}~\bibnamefont{Zhang}},
	\bibinfo{author}{\bibfnamefont{X.}~\bibnamefont{Dai}},
	\bibinfo{author}{\bibfnamefont{Z.}~\bibnamefont{Fang}}, \bibnamefont{and}
	\bibinfo{author}{\bibfnamefont{S.-C.} \bibnamefont{Zhang}},
	\bibinfo{journal}{Phys. Rev. B} \textbf{\bibinfo{volume}{82}},
	\bibinfo{pages}{045122} (\bibinfo{year}{2010}).
	
	\bibitem[{\citenamefont{Moore}(2010)}]{moore2010birth}
	\bibinfo{author}{\bibfnamefont{J.~E.} \bibnamefont{Moore}},
	\bibinfo{journal}{Nature} \textbf{\bibinfo{volume}{464}},
	\bibinfo{pages}{194} (\bibinfo{year}{2010}).
	
	\bibitem[{\citenamefont{Zhang et~al.}(2009)\citenamefont{Zhang, Liu, Qi, Dai,
			Fang, and Zhang}}]{Zhang2009}
	\bibinfo{author}{\bibfnamefont{H.}~\bibnamefont{Zhang}},
	\bibinfo{author}{\bibfnamefont{C.-X.} \bibnamefont{Liu}},
	\bibinfo{author}{\bibfnamefont{X.-L.} \bibnamefont{Qi}},
	\bibinfo{author}{\bibfnamefont{X.}~\bibnamefont{Dai}},
	\bibinfo{author}{\bibfnamefont{Z.}~\bibnamefont{Fang}}, \bibnamefont{and}
	\bibinfo{author}{\bibfnamefont{S.-C.} \bibnamefont{Zhang}},
	\bibinfo{journal}{Nature physics} \textbf{\bibinfo{volume}{5}},
	\bibinfo{pages}{438} (\bibinfo{year}{2009}).
	
	\bibitem[{\citenamefont{Fu}(2009)}]{Bi2Te3_1}
	\bibinfo{author}{\bibfnamefont{L.}~\bibnamefont{Fu}}, \bibinfo{journal}{Phys.
		Rev. Lett.} \textbf{\bibinfo{volume}{103}}, \bibinfo{pages}{266801}
	(\bibinfo{year}{2009}).
	
	\bibitem[{\citenamefont{Chen et~al.}(2009)\citenamefont{Chen, Analytis, Chu,
			Liu, Mo, Qi, Zhang, Lu, Dai, Fang et~al.}}]{Bi2Te3_2}
	\bibinfo{author}{\bibfnamefont{Y.}~\bibnamefont{Chen}},
	\bibinfo{author}{\bibfnamefont{J.~G.} \bibnamefont{Analytis}},
	\bibinfo{author}{\bibfnamefont{J.-H.} \bibnamefont{Chu}},
	\bibinfo{author}{\bibfnamefont{Z.}~\bibnamefont{Liu}},
	\bibinfo{author}{\bibfnamefont{S.-K.} \bibnamefont{Mo}},
	\bibinfo{author}{\bibfnamefont{X.-L.} \bibnamefont{Qi}},
	\bibinfo{author}{\bibfnamefont{H.}~\bibnamefont{Zhang}},
	\bibinfo{author}{\bibfnamefont{D.}~\bibnamefont{Lu}},
	\bibinfo{author}{\bibfnamefont{X.}~\bibnamefont{Dai}},
	\bibinfo{author}{\bibfnamefont{Z.}~\bibnamefont{Fang}}, \bibnamefont{et~al.},
	\bibinfo{journal}{Science} \textbf{\bibinfo{volume}{325}},
	\bibinfo{pages}{178} (\bibinfo{year}{2009}).
	
	\bibitem[{\citenamefont{Young and Kane}(2015)}]{dirac_semi}
	\bibinfo{author}{\bibfnamefont{S.~M.} \bibnamefont{Young}} \bibnamefont{and}
	\bibinfo{author}{\bibfnamefont{C.~L.} \bibnamefont{Kane}},
	\bibinfo{journal}{Physical review letters} \textbf{\bibinfo{volume}{115}},
	\bibinfo{pages}{126803} (\bibinfo{year}{2015}).
	
	\bibitem[{\citenamefont{Zhai}(2021)}]{Zhai_2021}
	\bibinfo{author}{\bibfnamefont{H.}~\bibnamefont{Zhai}},
	\emph{\bibinfo{title}{Ultracold Atomic Physics}}
	(\bibinfo{publisher}{Cambridge University Press}, \bibinfo{year}{2021}).
	
	\bibitem[{\citenamefont{Meir and Wingreen}(1992)}]{meir1992landauer}
	\bibinfo{author}{\bibfnamefont{Y.}~\bibnamefont{Meir}} \bibnamefont{and}
	\bibinfo{author}{\bibfnamefont{N.~S.} \bibnamefont{Wingreen}},
	\bibinfo{journal}{Phys. Rev. Lett.} \textbf{\bibinfo{volume}{68}},
	\bibinfo{pages}{2512} (\bibinfo{year}{1992}).
	
	\bibitem[{\citenamefont{Haug et~al.}(2008)\citenamefont{Haug, Jauho
			et~al.}}]{haug2008quantum}
	\bibinfo{author}{\bibfnamefont{H.}~\bibnamefont{Haug}},
	\bibinfo{author}{\bibfnamefont{A.-P.} \bibnamefont{Jauho}},
	\bibnamefont{et~al.}, \emph{\bibinfo{title}{Quantum kinetics in transport and
			optics of semiconductors}}, vol.~\bibinfo{volume}{2}
	(\bibinfo{publisher}{Springer}, \bibinfo{year}{2008}).
	
	\bibitem[{\citenamefont{Datta}(1997)}]{datta1997electronic}
	\bibinfo{author}{\bibfnamefont{S.}~\bibnamefont{Datta}},
	\emph{\bibinfo{title}{Electronic transport in mesoscopic systems}}
	(\bibinfo{publisher}{Cambridge university press}, \bibinfo{year}{1997}).
	
	\bibitem[{\citenamefont{Bruus and Flensberg}(2004)}]{Bruus&Flensburg}
	\bibinfo{author}{\bibfnamefont{H.}~\bibnamefont{Bruus}} \bibnamefont{and}
	\bibinfo{author}{\bibfnamefont{K.}~\bibnamefont{Flensberg}},
	\emph{\bibinfo{title}{Many-body quantum theory in condensed matter physics:
			an introduction}} (\bibinfo{publisher}{OUP Oxford}, \bibinfo{year}{2004}).
	
	\bibitem[{\citenamefont{Sandvik}(2010)}]{sandvik2010computational}
	\bibinfo{author}{\bibfnamefont{A.~W.} \bibnamefont{Sandvik}}, in
	\emph{\bibinfo{booktitle}{AIP Conference Proceedings}}
	(\bibinfo{organization}{American Institute of Physics},
	\bibinfo{year}{2010}), vol. \bibinfo{volume}{1297}, pp.
	\bibinfo{pages}{135--338}.
	
	\bibitem[{\citenamefont{Joe et~al.}(2006)\citenamefont{Joe, Satanin, and
			Kim}}]{joe2006classical}
	\bibinfo{author}{\bibfnamefont{Y.~S.} \bibnamefont{Joe}},
	\bibinfo{author}{\bibfnamefont{A.~M.} \bibnamefont{Satanin}},
	\bibnamefont{and} \bibinfo{author}{\bibfnamefont{C.~S.} \bibnamefont{Kim}},
	\bibinfo{journal}{Physica Scripta} \textbf{\bibinfo{volume}{74}},
	\bibinfo{pages}{259} (\bibinfo{year}{2006}).
	
	\bibitem[{\citenamefont{Miroshnichenko
			et~al.}(2010)\citenamefont{Miroshnichenko, Flach, and Kivshar}}]{FANO_RMP}
	\bibinfo{author}{\bibfnamefont{A.~E.} \bibnamefont{Miroshnichenko}},
	\bibinfo{author}{\bibfnamefont{S.}~\bibnamefont{Flach}}, \bibnamefont{and}
	\bibinfo{author}{\bibfnamefont{Y.~S.} \bibnamefont{Kivshar}},
	\bibinfo{journal}{Rev. Mod. Phys.} \textbf{\bibinfo{volume}{82}},
	\bibinfo{pages}{2257} (\bibinfo{year}{2010}).
	
	\bibitem[{\citenamefont{Carbotte}(2016)}]{TiltDirac1}
	\bibinfo{author}{\bibfnamefont{J.}~\bibnamefont{Carbotte}},
	\bibinfo{journal}{Phys. Rev. B} \textbf{\bibinfo{volume}{94}},
	\bibinfo{pages}{165111} (\bibinfo{year}{2016}).
	
	\bibitem[{\citenamefont{Yang et~al.}(2018)\citenamefont{Yang, Wang, and
			Liu}}]{TiltDirac2}
	\bibinfo{author}{\bibfnamefont{Z.-K.} \bibnamefont{Yang}},
	\bibinfo{author}{\bibfnamefont{J.-R.} \bibnamefont{Wang}}, \bibnamefont{and}
	\bibinfo{author}{\bibfnamefont{G.-Z.} \bibnamefont{Liu}},
	\bibinfo{journal}{Phys. Rev. B} \textbf{\bibinfo{volume}{98}},
	\bibinfo{pages}{195123} (\bibinfo{year}{2018}).
	
	\bibitem[{\citenamefont{Zheng et~al.}(2020)\citenamefont{Zheng, Duan, Wang, Li,
			Deng, and Wang}}]{TiltDirac3}
	\bibinfo{author}{\bibfnamefont{S.-H.} \bibnamefont{Zheng}},
	\bibinfo{author}{\bibfnamefont{H.-J.} \bibnamefont{Duan}},
	\bibinfo{author}{\bibfnamefont{J.-K.} \bibnamefont{Wang}},
	\bibinfo{author}{\bibfnamefont{J.-Y.} \bibnamefont{Li}},
	\bibinfo{author}{\bibfnamefont{M.-X.} \bibnamefont{Deng}}, \bibnamefont{and}
	\bibinfo{author}{\bibfnamefont{R.-Q.} \bibnamefont{Wang}},
	\bibinfo{journal}{Phys. Rev. B} \textbf{\bibinfo{volume}{101}},
	\bibinfo{pages}{041408} (\bibinfo{year}{2020}).
	
\end{thebibliography}
\end{document}